\documentclass[aps,prb,twocolumn,floatfix,superscriptaddress,amsmath,amssymb]{revtex4-2}

\usepackage{epsfig}
\usepackage{graphicx}
\usepackage{dcolumn}
\usepackage{bm}
\usepackage{multirow}
\usepackage{CJK}
\usepackage{float}
\usepackage[pdftex]{color}
\usepackage[dvipsnames]{xcolor}
\usepackage{physics}

\begin{document}
\title[]{Universal entanglement correction induced by relevant deformations at the quantum critical point}
\date{\today}

\newcommand{\ugent}[0]{Department of Physics and Astronomy, University of Ghent, Belgium}

\newcommand{\ucas}[0]{Kavli Institute for Theoretical Sciences, University of Chinese Academy of Sciences, Beijing 100190, China}

\author{Rui-Zhen Huang}
\affiliation{\ugent}

\author{Chen Peng}
\email{pengchen@ucas.ac.cn}
\affiliation{\ucas}

\begin{abstract}
    Local relevant deformations are important tool to study universal properties of quantum critical points. We investigate the effect of small relevant deformations on the bi-partite entanglement entropy at the quantum critical points. Within the quantum critical region, a universal power-law correction in the entanglement entropy induced by the relevant operator is found in both one- and two-dimensional critical lattice models. The exponent of the power-law correction term is determined by the scaling dimension of the relevant operator. Based on numerical simulations and scaling theory argument, it is conjectured that such a universal power-law correction in the entanglement entropy is universal for Lorentz invariant quantum critical points. Without Lorentz invariance, it is found the exponent in the power-law correction term does not fit in with the scaling argument in models with a dynamical exponent $z=2$ in two dimension. This may be because the relevant operator added in the lattice model corresponds to complicated operators in the corresponding conformal field theory. Our study provides a different perspective to extract universal information of quantum critical points. We expect it would be useful to detect unique properties of topological quantum phase transitions. 
\end{abstract}
\maketitle

\section{Introduction}
Along with the recent fast development of quantum computing technologies, the study and understanding of quantum entanglement becomes one of the most important topics. As a key concept in quantum theories, the entanglement has already inspired many researches for both high energy and condensed matter physics in recent years. The measure of the amount of quantum entanglement, the entanglement entropy (EE) $S := - \mathrm{tr} \left( \rho_A \log(\rho_A) \right)$ ($\rho_A$ is the reduced density matrix for the sub-region $A$ in the whole system) deepens our understanding of quantum matter in many aspects~\cite{hastings2007area,Eisert2010,nishioka2018entanglement,Zeng2019_book,peps_2004,frank_2006,mps_classification2011,chen_spt_2013, Levin2006, Kitaev2006, mps_peps_review2021,entanglement_qcp1,entanglement_qcp2,entanglement_qcp3,entanglement_qcp4, corner_EE1,corner_EE1,corner_EE2,corner_EE3,corner_EE4,corner_EE5,corner_EE6,ent_dir_zhu}, especially in the study of gapped quantum phases. The famous area law for the EE in gapped ground states not only reveals the relation between degrees of freedom in a quantum system~\cite{hastings2007area,Eisert2010,nishioka2018entanglement}, but also inspired novel quantum many-body frameworks~\cite{Zeng2019_book,peps_2004,frank_2006}.Moreover, a finer structure of the entanglement, including the degeneracy in the entanglement spectrum and the long-range entanglement, becomes the key to classify gapped topological quantum phases~\cite{Zeng2019_book,mps_classification2011,chen_spt_2013, Levin2006, Kitaev2006, mps_peps_review2021}. 

The entanglement scaling is also found important applications in critical systems~\cite{entanglement_qcp1,entanglement_qcp2,entanglement_qcp3,entanglement_qcp4, corner_EE1,corner_EE1,corner_EE2,corner_EE3,corner_EE4,corner_EE5,corner_EE6,ent_dir_zhu,dqcp_corner_EE1,dqcp_corner_EE2,entanglement_qcp4,Cho2017,huang2024emergent}. The logarithmic scaling of the EE becomes one of the canonical methods to identify one-dimensional quantum critical points and extract the central charge for the underlying conformal field theory (CFT)~\cite{entanglement_qcp1,entanglement_qcp2,entanglement_qcp3,entanglement_qcp4,nishioka2018entanglement}. For two-dimensional quantum critical points, the corner EE is found to be related to Goldstone modes and is utilized to diagnose a class of controversial quantum phase transitions~\cite{dqcp_corner_EE1,dqcp_corner_EE2}. Despite these successful applications of entanglement in criticality, the information of the fixed point is not included there. For example, one can extract the central charge from the scaling the EE, however there may be different CFTs with the same central charge~\cite{francesco2012conformal, Blumenhagen:2009zz}. A natural question is thus how one can find universal data from the EE. Recently it is found in one dimensions, the entanglement spectrum can be used to detect the operator contents of the corresponding CFT in some cases, provided with some other input information~\cite{Laeuchli2013,entanglement_qcp4,Cho2017,huang2024emergent}. For two-dimensional universalities, or even more general cases without emergent Lorentz invariance, it is still an open question. 

From the scaling theory viewpoint, the universal scaling form of the EE~\cite{nishioka2018entanglement} at the quantum critical point allows one to resemble local physical quantities and regard it as a quantity with a scaling dimension. For two-dimensional quantum critical points with emergent Lorentz invariance, such as the two-dimensional critical quantum Ising model on the square lattice, the EE satisfies an area law. One may expect close to the quantum critical point, a scaling theory may be written down for the EE and its deviation from the area law be universal. There has already been research in this area, primarily focusing on conformal field theory in (1+1)D and free field theories~\cite{Cardy_2010,Calabrese_2010}. For lattice models in the quantum critical region, especially two-dimensional models, there is still much to explore.

In this paper, we study the effect of small relevant deformations on the entangle for both one- and two-dimensional quantum critical points with emergent Lorentz invariance on the lattice. In one dimension, we use the finite-entanglement scaling approach based on the infinite matrix product state (MPS) representation~\cite{white1992density,Pollmann2009,vmps1,mps_peps_review2021}. The finite length scale $\xi$ in the MPS is introduced by the entanglement cutoff. In two dimensions, we use the correlation matrix technique~\cite{Peschel_2009_cm} to study the EE of free Dirac fermions. We consider relevant deformations with or without $U(1)$ symmetries to the critical fermions. Combing the numerical calculations and the scaling theory argument, we propose there is a universal power-law correction in the EE induced by the relevant deformation. After that we present our study of critical points without Lorentz invariance. Finally we give a summary and conclusion.

\begin{figure}[tbp]
\centering
\includegraphics[width=\columnwidth]{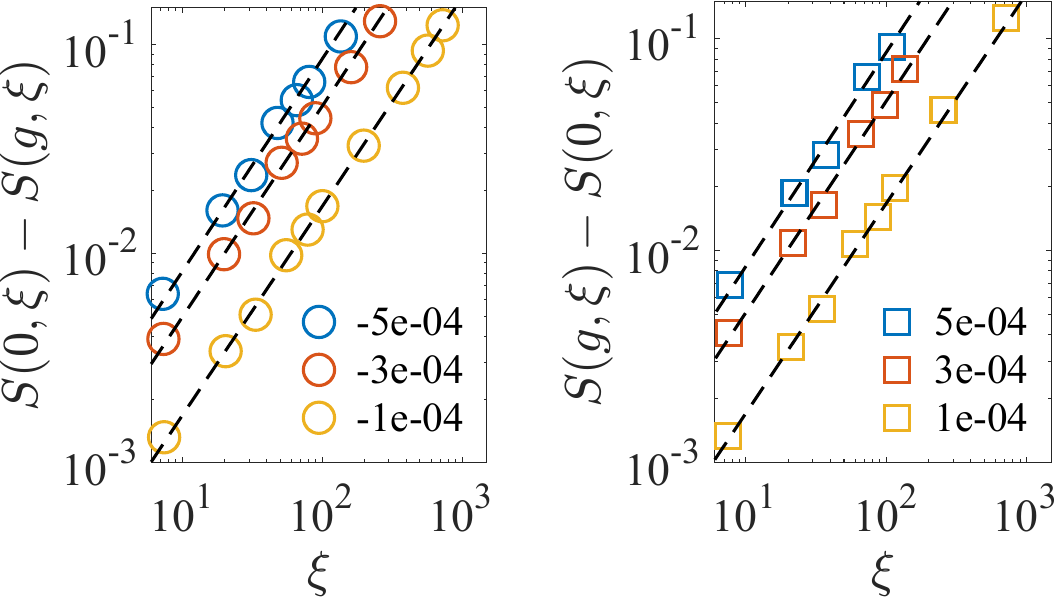}
  \caption{Power-law correction of the EE for the critical quantum Ising chain deformed by the transverse field $g = h - h_c$ obtained from the infinite uniform MPS. The exponent from the power-law fitting (dashed lines) are found to be $\Delta_g = 1.00$.}
\label{fig_ising}
\end{figure}

\begin{figure}[tbp]
\centering
\includegraphics[width=0.38\textwidth=]{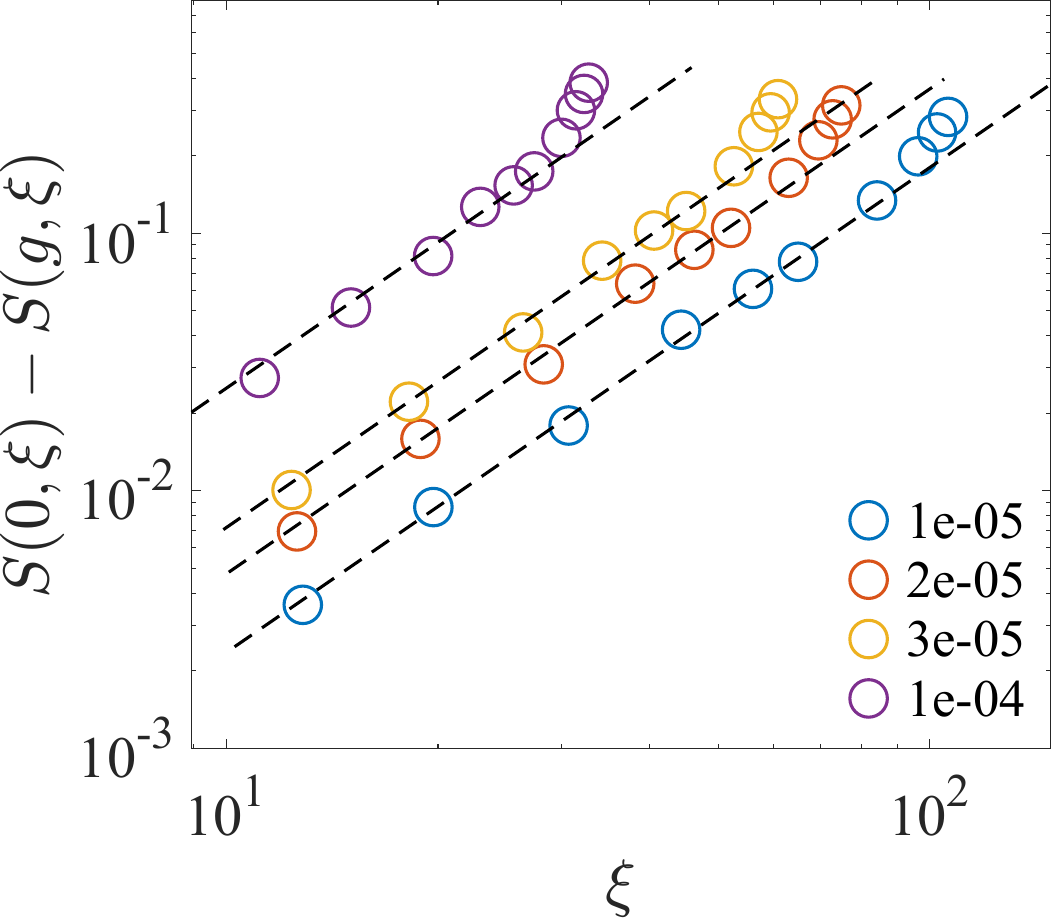}
  \caption{Power-law correction of the EE for the critical quantum Ising chain deformed by a longitudinal field $g=h_z$ obtained from the infinite uniform MPS. The exponent from the power-law fitting (dashed lines) is found to be $\Delta_g = 1.88$.}
\label{fig_ising2}
\end{figure}

\section{Entanglement correction in one dimensional quantum critical points}
Before we present our study of the entanglement properties of (2+1)D theories, we investigate the influence of relevant deformations in one-dimensional quantum critical points with a underlying (1+1)D CFT description. Since we are going to use the infinite uniform matrix product state (MPS) framework.~\cite{white1992density,mps_tebd,vmps1} to study the EE, the length scale therein is the correlation length $\xi$ induced by the entanglement cutoff. MPS is a generic class of tensor network states describing low-energy states of one-dimensional lattice models well. The bipartite EE for critical the ground states is known to scale logarithmically with respect to the finite length scale of the system $ S \sim \frac{c}{6} \log (\xi) $ where the sub-region $A$ is assumed to the left half spin chain. We add a relevant operator $H^\prime := g\,O_g$ to the Hamiltonian $H_0$ at the critical point and study the deviation of the logarithmic scaling for the EE. Resembling the EE as a conventional physical quantity (dimensionless), we expect the leading correction in the EE may be of the form 
\begin{equation}
    \delta S \sim g\,\xi ^{d+z-\Delta_g}    
    \label{eq_ds_1d}
\end{equation}
where $d=1$ is the space dimension, $z=1$ denotes the dynamical exponent, and $\Delta_g$ is the scaling dimension of $O_g$.  

We considered the quantum Ising chain 
\begin{equation}
    H = \sum_i \left( -Z_i Z_{i+1} - h X_i\right)
\end{equation}
where $Z$ and $X$ are Pauli matrices defined on the two-dimensional Hilbert space. The model has a global $Z_2$ symmetry. There is a quantum phase transition in the ground state at $h_c=1$ separating the symmetry-breaking phase $h<1$ and the symmetric phase when $h>1$. At the low energy, the critical Ising model is described by the A-series minimal model $M(4,3)$ CFT~\cite{francesco2012conformal}. There are two non-trivial relevant operator for the Ising CFT, including the spin and the energy operator. On the lattice they are realized by adding a longitudinal field $h_z$ or deforming the transverse field to $h_c+\delta h$, respectively. 

In Fig.~\ref{fig_ising}, we show the correction for the entanglement against the finite correlation length induced by adding a small relevant field $g = \delta h$. The relevant deformation is the energy operator $\epsilon$ with a scaling dimension $\Delta_\epsilon = 1$. The correction of the EE shows a perfect power-law behavior $\xi^p$ with an exponent $p \sim 1.00$. This is consistent with the form Eq.~\ref{eq_ds_1d}. Interestingly, we notice that there is a sign change for the EE correction term when adding a positive $\delta h > 0$ or a negative $\delta h < 0$ deformation. This is because the phase transition in the Ising model is induced by the energy operator and there is a shift for the critical point at finite length scales. Hence there exists an imbalance distributions for the EE and other physical quantities within the quantum critical region. 

We also consider deforming the critical point with a symmetry-breaking field $H^\prime = h_z \sum_i Z_i$. This lattice operator realizes the spin operator $\sigma$ with a scaling dimension $\Delta_\sigma = 1/8$ in the Ising CFT. The numerical results in Fig.~\ref{fig_ising2} confirm the power-law behavior and the exponent $p\sim 1.88$ is close to the exponent $d+z-\Delta_\sigma = 15/8$. 

The power-law correction for the EE found in the numerical calculation in Figs.~\ref{fig_ising} and \ref{fig_ising2} may be explained from the scaling theory argument. One can regard the EE or exponential of the EE as a physical quantity with a dimension $\Delta=0$ or $\Delta = -1$, respectively. The EE close to the critical point can be written down as
\begin{equation}
    S(g,\xi) \sim \frac{c}{6} \log \left( \xi \, f_1(g\,\xi^{d+z-\Delta_g})\right),
\end{equation}
where the finite length scale $\xi$ can be replaced by other scales in different finite scaling approaches. Assuming a small parameter $g$ and the scaling function being smooth, one can write down an approximated scaling theory up to the leading correction 
\begin{equation}
    S(g,\xi) \sim \frac{c}{6}\log (\xi) + c_g\, g \xi^{d+z-\Delta_g} + O(g^2)
\end{equation}
where $c_g$ is obtained from the expansion of the scaling function and assumed to be non-zero. It predicts a universal power-law correction for the EE with an exponent $p = d+z-\Delta_g$. This is consistent with our numerical results in Fig.~\ref{fig_ising} by deforming the energy and the spin operators.

\section{Entanglement entropy correction in two-dimensional free lattice fermions}
In two dimensions, the underlying (2+1)D CFT~\cite{francesco2012conformal} for the critical lattice models is not as powerful as the one-dimensional case. The numerical study of the EE is also challenging for, e.g. tensor networks or quantum Monte Carlo methods. Here we study the critical free lattice fermion models, which are well under control numerically. Thanks to the Wick theorem, the EE can be easily obtained from the single-particle correlation function~\cite{Peschel_2009_cm}. Note that since the fermion annihilation/creation operator has a scaling dimension $\Delta=1$, the relevant operators we can write down must be bi-linear fermion operators with an even fermion parity.

\subsection{\label{sub2}Entanglement entropy in free Dirac fermions}
In this section, we study the entanglement in critical free lattice fermion under small relevant perturbations. The critical lattice model is described by the Dirac fermion in the low energy. We consider $L\times L$ finite lattices with periodic boundary conditions on both $x$ and $y$ spatial directions. Two sub-regions $A$ of different geometries are considered, either a $\frac{L}{2} \times L$ cylinder without corners or a $\frac{L}{2} \times \frac{L}{2}$ square with four corners. The EE between the sub-region $A$ and its rest part $\bar{A}$ is calculated from the single-particle correlation defined within the sub-region~\cite{Peschel_2009_cm}. 

\begin{figure}[t]
    \centering
    \includegraphics[width=0.5\textwidth]{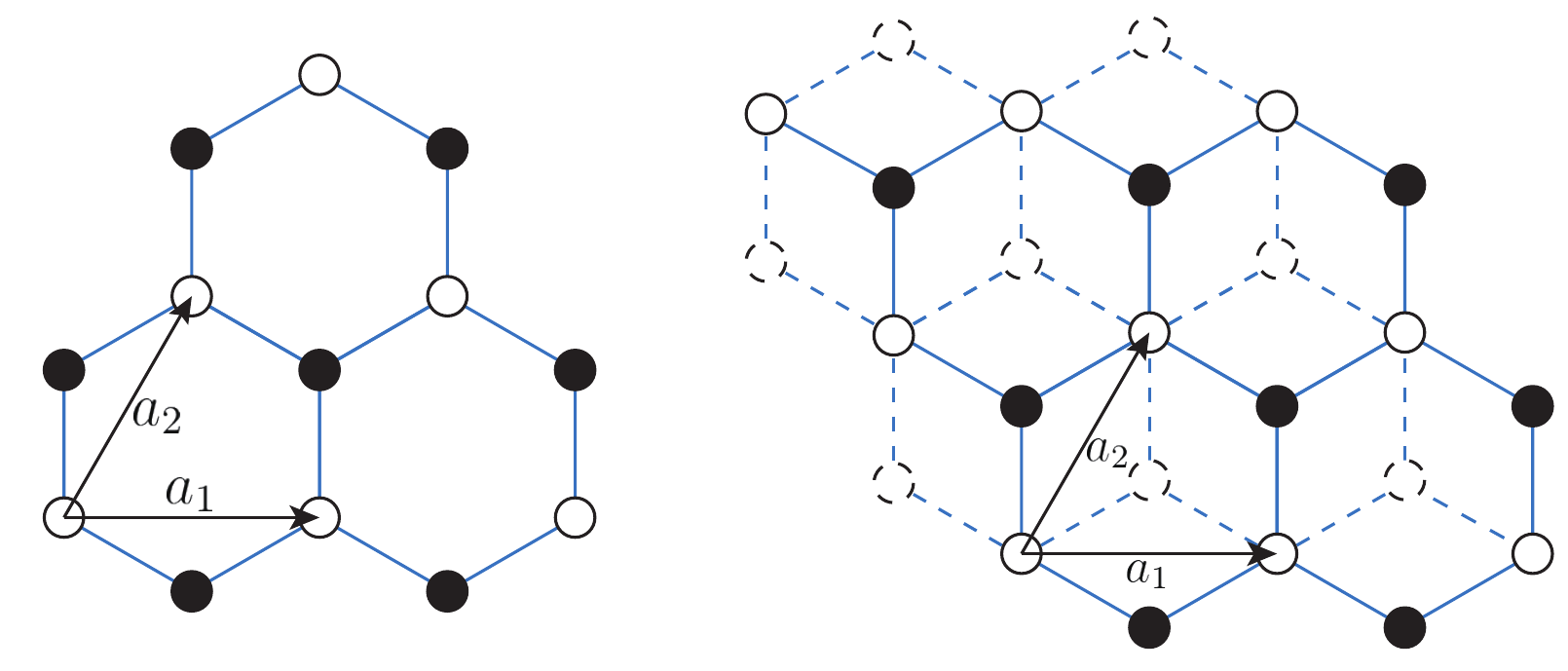}
    \caption{(a) The honeycomb lattice studied in Sec.~\ref{sub2}. (b) The bernal-stacked bilayer graphene studied in Sec.~\ref{secC}. The upper and lower layer are shown with solid and dash lines, respectively. The inter-layer hopping $t_i$ connects different sub-site A and B of each layer.}
    \label{fig:lattice}
\end{figure}

We consider the simplest lattice model that realizes free Dirac fermions on the honeycomb lattice as shown in Fig.~\ref{fig:lattice} (a) 
\begin{equation}
    H = H_0 + H^\prime    
\end{equation}
where $H_0$ denotes the critical free fermion term
\begin{equation}
{H_0} =  - t\sum\limits_{ < ij >,\sigma } {\left( {c_{i,\sigma}^\dag {c_{j,\sigma}} + c_{j,\sigma}^\dag {c_{i,\sigma}}} \right)},
\label{eq:ham}
\end{equation}
and $H^\prime = g\,O_g$ denotes a small relevant operator. The gapless point $g=0$ can be viewed as a fine-tuning quantum critical point between the $g>0$ and $g<0$ gapped phases. At the gapless point, there are two Dirac points at the $K$ and $K^\prime$ points in the first Brillouin zone. The number of Dirac points must be even for lattice models~\cite{graphene_review_2009}.  

Without a finite fermi surface, the EE in the Dirac fermion model satisfies an area law $S \sim \alpha L$. Similar to the one-dimensional case, regarding the EE as a conventional physical quantity, a power-law correction
\begin{equation}
    \delta S \sim g L^{d+z+1-\Delta_g}
    \label{Eq_dEE2}
\end{equation}
is expected. Interestingly, the gapped phases induced by a relevant field $g$ also satisfy an area law $S \sim \alpha_g L$ for the EE at large length scales $L \gg \xi_g$. One notices that it is very different from the the one-dimensional case studied in the last section, in which the EE satisfies different scaling forms between the critical point and the gapped phase. This makes the identification of the EE response to a mall relevant field more challenging numerically in two dimensions. Nevertheless, we expect that within the quantum critical region $L \ll \xi_g$, the correction in the EE induced by the small relevant operator can be observed, assuming the area law coefficient does not change much for a small deformation.

\subsubsection{The EE correction induced by a small mass deformation}
We investigate the scaling behavior of the EE in the vicinity of the Dirac fermion with a small deformation -- a $U(1)$ symmetric mass term $m$ 
\begin{equation}
    H^\prime = m\sum\limits_{i,\sigma} {{{( - 1)}^i}c_{i,\sigma}^\dag {c_{i,\sigma}}} 
    \label{eq:mass}
\end{equation}
which stands for the on-site staggered chemical potential. $H^\prime$ opens a small energy gap $1/\xi_m$ at the Dirac points. When $m$ changes sign, an inversion of energy bands occur. The quantum critical region is well-defined for small systems $L \ll \xi_m \sim 1/\vert m \vert$. The positive and negative $m$ regions are related by a unitary transformation, therefor we only consider $m>0$ in the following.

\begin{figure}[t]
    \centering
    \includegraphics[width=0.38\textwidth]{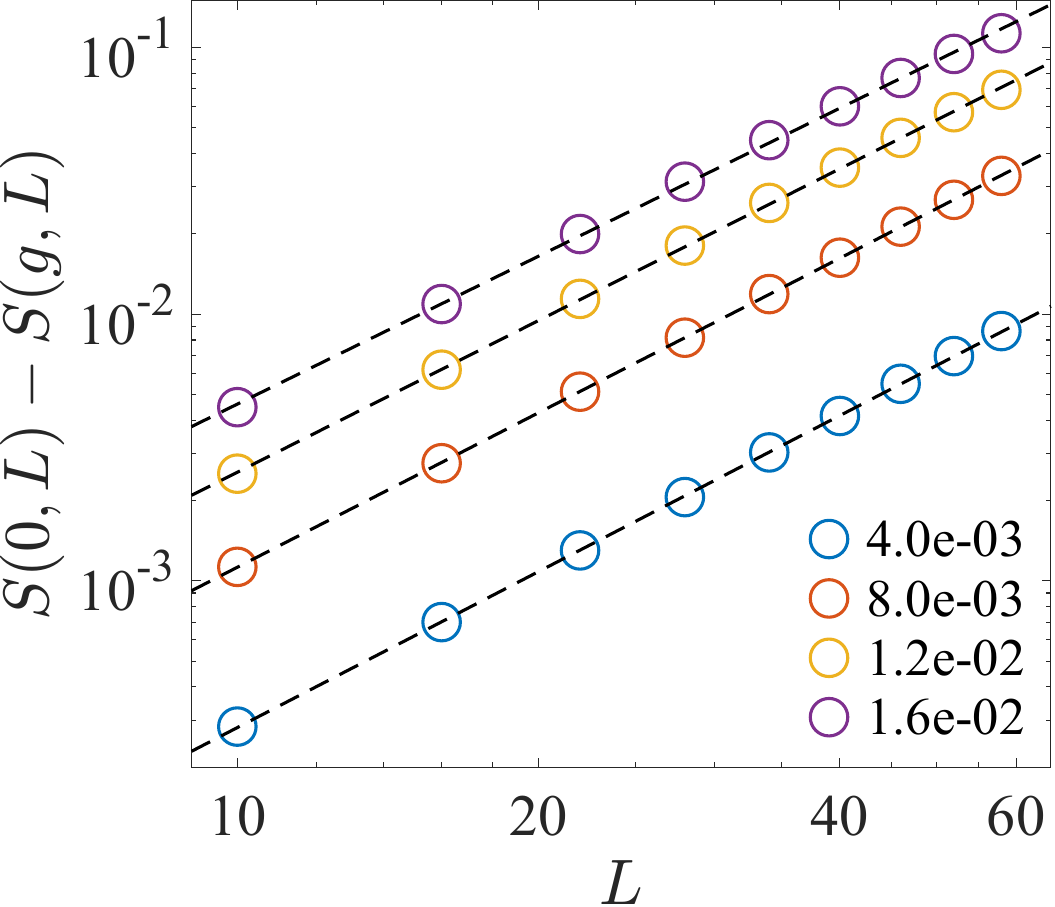}
    \caption{Power-law correction of the EE induced by a small mass term to the critical free fermion on the Honeycomb lattice. The exponent is found to be 1.94 from the power-law fitting (dashed line). The sub-region is of a cylinder geometry.}
    \label{fig-SvN_hex_torus}
\end{figure}

We first consider the sub-region $A$ of a cylinder geometry with a smooth entangling surface. The area law term should dominate the scaling for the EE. As expected, the EE shows perfect area law in the quantum critical region. Figure.~\ref{fig-SvN_hex_torus} shows the scaling behavior of the leading correction 
\begin{equation}
    S^{(1)}(g,L) = S(g,L) - S(0,L)   
    \label{Eq_dEE}
\end{equation}
which satisfies a perfect power-law fitting. The exponent found $1.94$ is close to $2$, consistent with the power-law form $L^{d+z+1-\Delta_m}$ in which the mass term has a scaling dimension $\Delta_m = 2$ for the free Dirac fermion. 

\begin{figure}[t]
    \centering
    \includegraphics[width=0.5\textwidth]{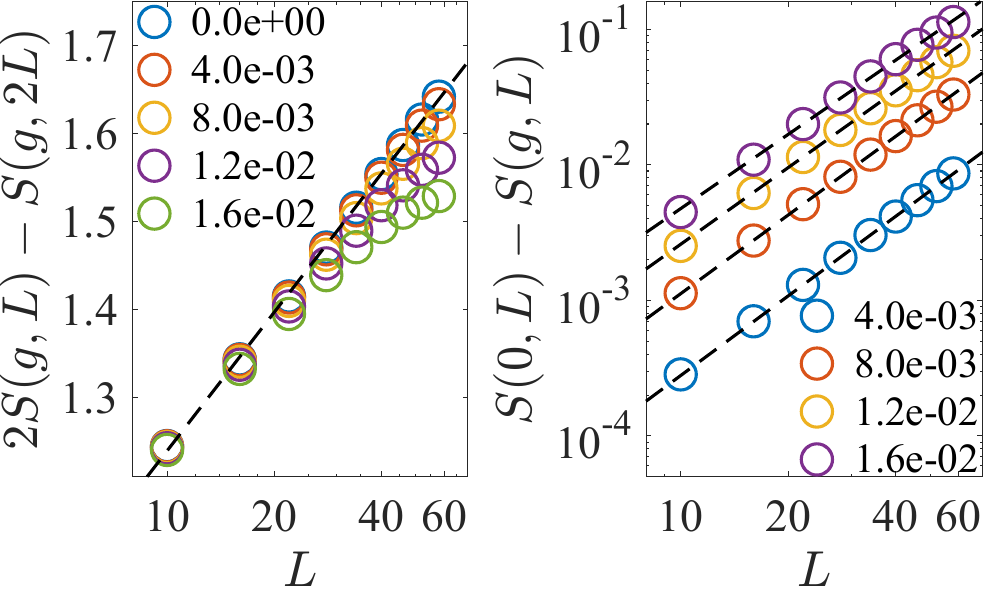}
    \caption{The EE for the free fermion on the Honeycomb lattice with a sub-region of the square geometry. (a) Logarithmic scaling term $S^c$ in the EE caused by the corner in the sub-region $A$. (b) Power-law correction of the EE induced by a small mass term. The exponent is found to be 1.95 from the power-law fitting (dashed line).}
    \label{fig:SvN-dev-hex-corner}
\end{figure}

Now we begin to study the sub-region $A$ of a square geometry with corners. As analysed above, the existence of singularity points in the entangling surface leads to another divergent logarithmic EE. To study the corner contribution, one needs to properly subtract the area law part. Here we define 
\begin{equation}
    S^c(g,L) = \vert S(g,2L) - 2 S(g,L) \vert
\end{equation}
by using the $2L \times 2L$ and $L \times L$ lattices~\cite{note_log}. Such a logarithmic divergent term is accurately produced in the finite size scaling, see Fig.~\ref{fig:SvN-dev-hex-corner} (a). Turing on a small mass perturbation, one can also observe small deviations from the logarithmic form at large lengths. This deviation signals the correction induced by the deformation. In Fig.~\ref{fig:SvN-dev-hex-corner} (b), we show the scaling of the correction $S^{(1)}(g,L)$ defined in Eq.~\ref{Eq_dEE}. It also satisfies a perfect power-law fitting with an exponent $1.95$ which is also consistent with the form $L^{d+z+1-\Delta_m}$. The existence of corners in the sub-region $A$ does not change the leading correction by the deformation. One may wonder the corner contribution also contains a correction term with a smaller exponent. It belongs to next leading contribution, which is beyond the scope of the present research.

\begin{figure}[b]
    \centering
    \includegraphics[width=0.38\textwidth]{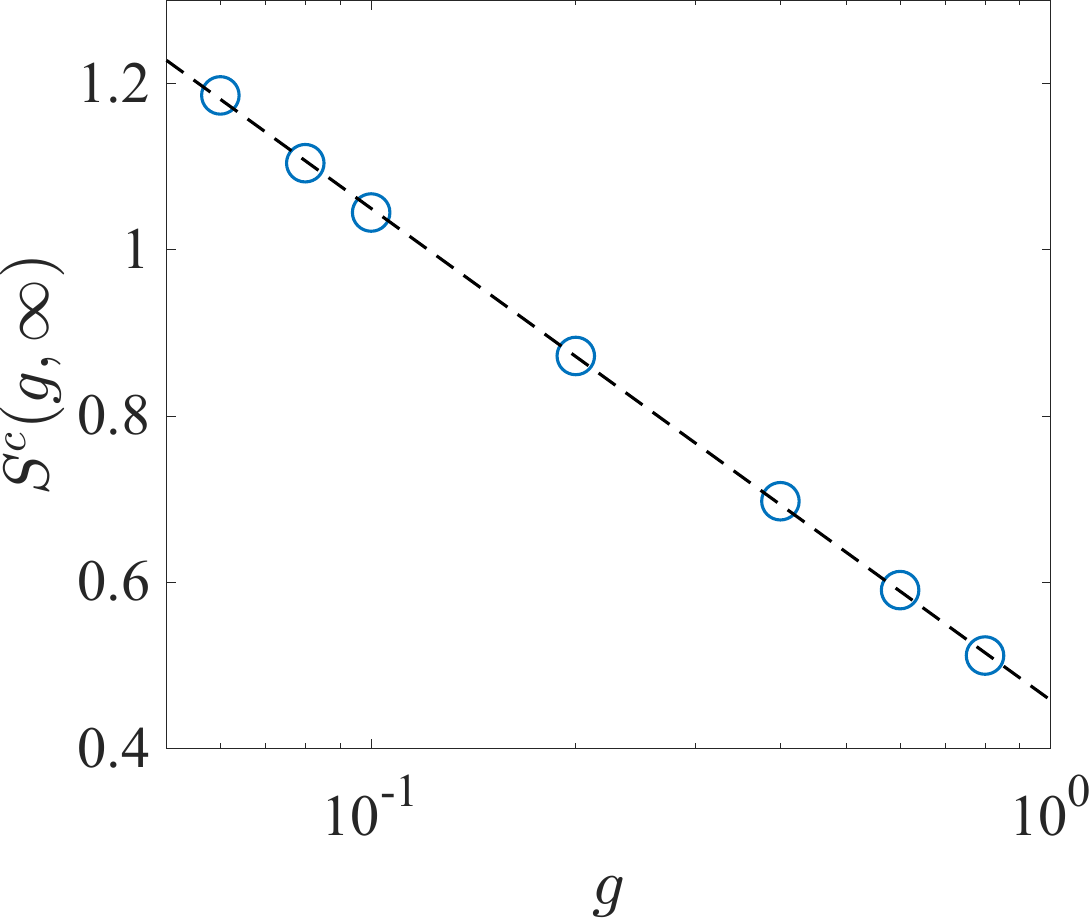}
    \caption{Finite mass scaling for the corner EE $S^c$. $S^c(g,\infty)$ is obtained by extrapolating the finite size $L$ to infinitely large in $S^c(g,L)$.}
    \label{fig:SvN-hex-converge}
\end{figure}

We have concentrated on the finite size scaling ($L\ll\xi$) approach so far. It is interesting that we can also use a different way to study the logarithmic term $S^c(g,L)$. For a given mass term, we first calculate the converged value at the thermodynamic limit $S^c(g, L\rightarrow \infty)$. The inverse mass becomes the length scale, which can be used to perform a finite mass scaling analysis. At the small mass regime, the corner EE should behave like $S^c(m) \sim \log(1/m)$. This is verified in Fig.~\ref{fig:SvN-hex-converge}, where the numerical data shows an accurate finite mass scaling behavior. 

\begin{figure}[t]
    \centering
    \includegraphics[width=0.38\textwidth]{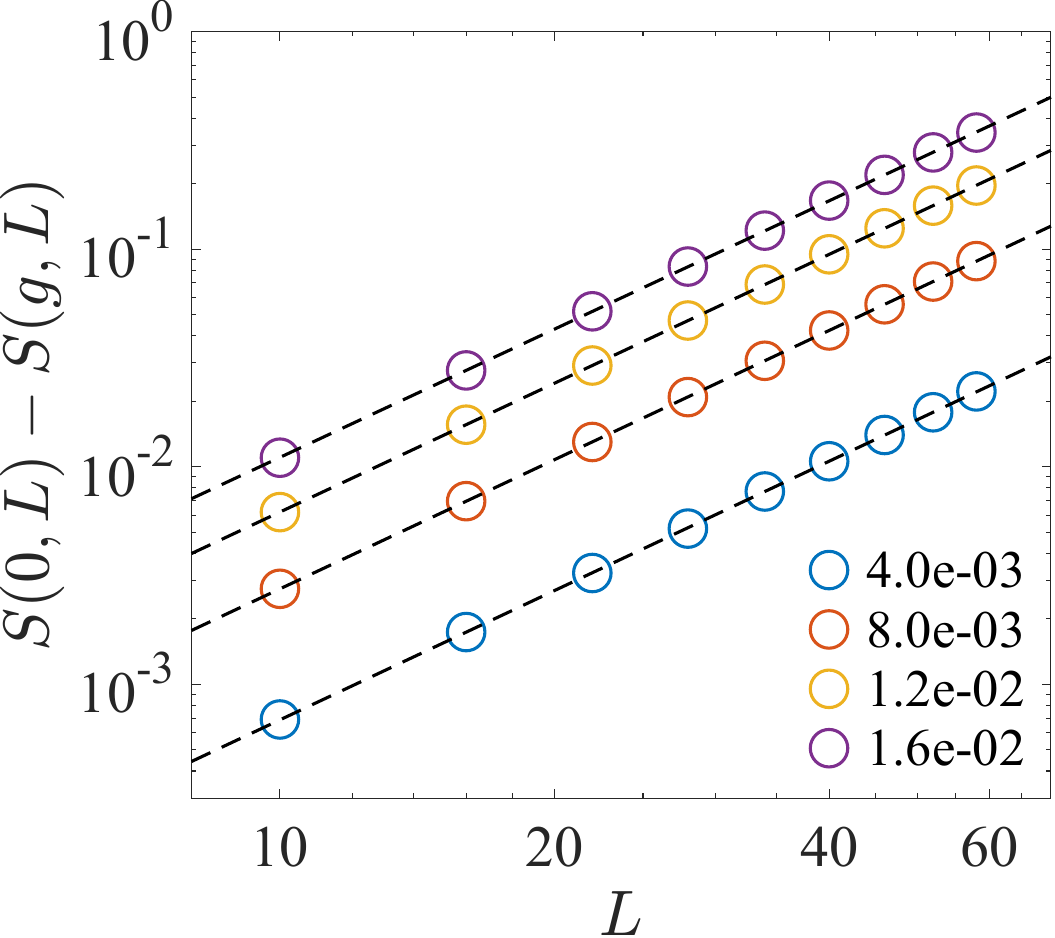}
    \caption{Power-law correction of the EE induced by a small pairing term to the critical free fermion on the Honeycomb lattice. The exponent is found to be 1.96 from the power-law fitting (dashed line). The sub-region is of a cylinder geometry.}
    \label{fig:SvN-dev-pair-cylinder}
\end{figure}

\subsubsection{The EE correction induced by a small pairing deformation}
In the previous subsection, we have shown that a $U(1)$ symmetric mass term could contribute to a universal power-law correction. Here we switch to an s-wave pairing (superconducting) deformation 
\begin{equation}
    H^\prime = g \sum\limits_{ i } {\left( {c_{i \uparrow }^\dag c_{i \downarrow }^\dag  + {c_{i \downarrow }}{c_{i \uparrow }}} \right)}. 
\label{eq:pair}
\end{equation}
where $g$ is small. It has the same scaling dimension as the mass term $\Delta_g = 2$. This superconducting term breaks the global $U(1)$ symmetry and results in a fully gapped superconductor ground state. The s-wave pairing ensures a small energy gap opening exactly at the Dirac points. When $g$ changes sign, the single particle energy band also inverses, in which the vacuum ground state becomes fully occupied state. The two phases are related by a canonical transformation. Note that there is no global $U(1)$ symmetry, the correlation matrix should also include the correlation function for the annihilation (creation) operators. 

We first study the EE correction for the sub-region of a cylinder geometry. As shown in Fig.\ref{fig:SvN-dev-pair-cylinder}, the correction term induced by a small pairing deformation shows a power-law behavior accurately. The exponent is found to be $1.96$, which is again consistent with the scaling form $L^{d+z+1-\Delta_g}$.

\begin{figure}[t]
    \centering
    \includegraphics[width=0.5\textwidth]{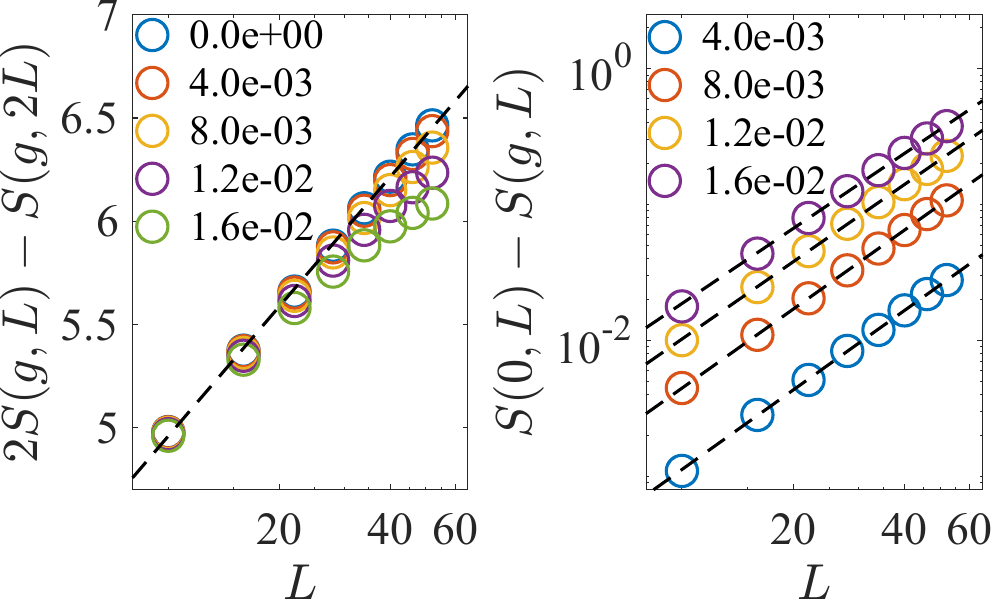}
    \caption{EE for the free fermion on the Honeycomb lattice with a sub-region of the square geometry. (a) The Logarithmic scaling term in the EE caused by the corner in the sub-region $A$. (b) The power-law correction of the EE induced by a small pairing term. The exponent is found to be 1.94 from the power-law fitting (dashed line).}
    \label{fig:SvN-hex-pair}
\end{figure}

Next we consider the sub-region of the square geometry. The numerical results for the $S_c$ is shown in Fig.~\ref{fig:SvN-hex-pair}. It shows a perfect logarithmic scaling behavior. Again, turning small pairing deformations, one can see there are deviations from the perfect logarithmic scaling. The correction in the EE induced by the deformation shows an accurately power-law behavior, see Fig.~\ref{fig:SvN-hex-pair} (b). The exponent is fitted to be $\delta_g \sim 1.94$ is again consistent with the scaling form Eq.~\ref{Eq_dEE2}.

\subsection{Scaling theory for the universal correction in (2+1)D quantum critical point}
Here we give a scaling theory argument to the power-law correction term found in the numerical calculations. In (2+1)D the EE for critical theories with conformal symmetry satisfy the area law for a sub-region with a smooth entangling surface. With a small relevant deformation $g\,\hat{O}_g$ to the critical point, we conjecture a scaling theory for the EE
\begin{equation}
    S(g,L) \sim L\, f \left(g\, L^{d+z-\Delta_g}\right)
\end{equation}
where $\Delta_g<d+z$ is the scaling dimension of the operator $O_g$. Similar to the one-dimensional case, assuming the scaling function $f$ to be smooth and a small enough perturbation $g$, one maybe expands the scaling theory in terms of $g$
\begin{equation}
    S(g,L) \sim S(0,L) + a_1\,g\,L^{d+z+1-\Delta_g} + O(g^2)
\label{Eq_EE_1}
\end{equation}
where $S(0,L) \sim a_0 L$ is the leading area law contribution. This gives rise to the universal power-law correction in Eq.~\ref{Eq_dEE2}. Such a universal correction can be detected when the length scale is much smaller than the correlation length, $L \ll \xi_g$. It should be noted that this leading correction relies on a non-zero expansion constant $a_1$. The power-law correction shown above may disappear and the leading correction comes from one order higher expansion $O(g^2)$ when $c_1$ disappears accidentally. 

The EE distribution may become more complicated if the entangling surface contains singular points. It is known that a singular corner in the sub-region contributes an extra logarithmic divergent term $\log (L)$ compared with those with a smooth entangling surface. Thus when the entangling surface contains corners, one may rewrite the EE as 
\begin{equation}
    \begin{aligned}
        S(g,L) = a_0 L + a_c \log(L) + a_1 g L^{d+z+1-\Delta_g} + O(g^2)
    \end{aligned} 
\label{Eq_EE_2}
\end{equation}
where only leading corrections are considered. There may exist a power-law correction term from the corner contribution, which has a smaller exponent, hence difficult to demonstrate in numerical simulations. 

Besides critical theories with emergent conformal symmetry, we are also interested in critical points without emergent Lorentz invariance, $z > 1$, at the low-energy limit. We expect the scaling theory, especially the EE correction in Eqs.~\ref{Eq_dEE} and \ref{Eq_dEE2} hold when the dynamical exponent is close to that in Lorentz invariant systems $z \sim 1$. 

\begin{figure}[tbp]
    \centering
    \includegraphics[width=0.5\textwidth]{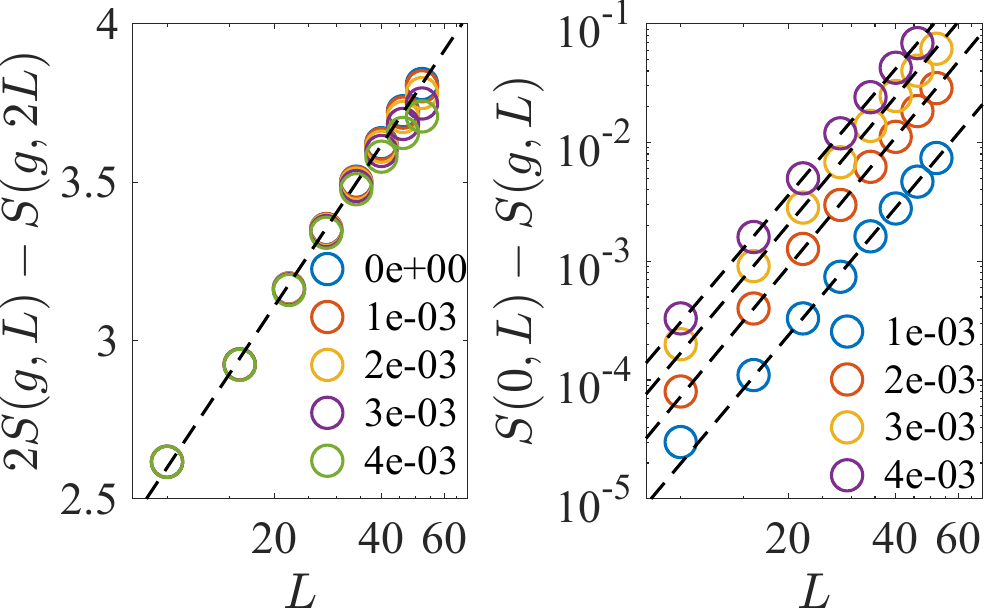}
    \caption{EE for the BSBG model ($t_i = 0.3$) with a sub-region of the square geometry. 
    (a) The Logarithmic scaling term in the EE caused by the corner in the sub-region $A$. (b) The power-law correction of the EE induced by a mass term. The exponent is found to be 3.6 from the power-law fitting (dashed line).}
    \label{fig:SvN-dev-BSBG}
\end{figure}

\subsection{\label{secC}Entanglement entropy in critical fermions without Lorentz invariance}
The critical lattice models studied above admit Lorentz invariance ($z=1$) at the low energy. From the scaling argument, this is not a necessary requirement in principle. For free fermions, a non-integer dynamical exponent is usually realized in models with long-range hoppings. For simplicity, here we study critical free lattice fermions with a dynamical exponent $z=2$, whose low-energy modes read $\omega \sim k^2$. Note that the scaling dimension of the fermionic annihilation operator $\psi$ does not rely on the value of $z$. A simple dimension counting gives $\Delta_\psi = 1$ for the free fermion in two dimensions. In the following, we are going to rely on numerical analysis to study the EE.

We consider the bernal-stacked bi-layer graphene~\ref{fig:lattice} with both intra- ($t=1$) and inter-layer ($t_i$) hoppings. The model admits two quadratic band touching points when $t_i \neq 0$, which results in a dynamical exponent $z=2$ at the low-energy. Deforming the critical model by adding a staggered mass term $m$ in both layers opens a small energy gap at the band-touching points. A simple dimensional counting tells that the scaling dimension of the relevant operator is $\Delta_m = 2$. The leading EE correction is thus expected to be the form of $\delta S \sim m L^3$.

\begin{figure}[t]
    \centering
    \includegraphics[width=0.38\textwidth]{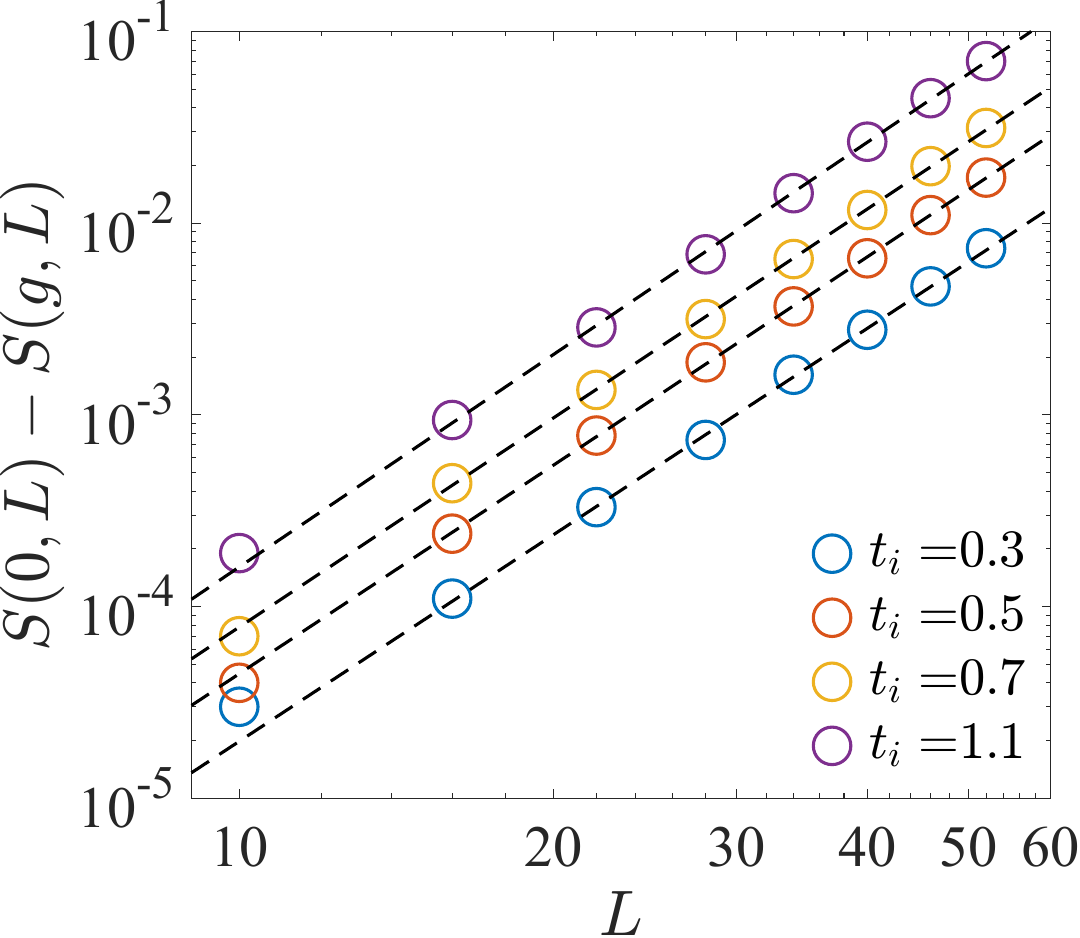}
    \caption{Power-law EE correction induced by a small mass term $m=0.001$ for different inter-layer hoppings.}
    \label{fig:SvN-dev-BSBG-diff-t}
\end{figure}

Figure.~\ref{fig:SvN-dev-BSBG} shows the EE for a given inter-layer hopping $t_i = 0.3$. It turns out the EE shows similar behavior as that in the Lorentz invariant theories. The corner contribution $S^c$ satisfies a logarithmic scaling for this model. Turning on the small mass deformation, it becomes deviated from the logarithmic form. Moreover, the leading correction induced by the small mass deformation is of a power-law form. However, the exponent is fitted to be $\sim 3.6$, which does not fit in with the scaling form $L^{d+z+1-\Delta_m}$. We also check the power-law correction for different inter-layer hopping $t_i$. Figure.~\ref{fig:SvN-dev-BSBG-diff-t} demonstrates that $S^c$ remains to be a power-law form with the same exponent. Considering the fact that, the space-time dimension and the dimension of the fermion operators are all integers, this is a surprising result. The scaling argument does not apply here. In a different model, the free fermion model on the square lattice with a flux~\cite{Zeng2018,hou2013}, we find the power-law correction due to the variation of the flux gives an power-law correction for the EE, which again can not be explained by the scaling argument. 

These results indicate that, small relevant deformations indeed induce a power-law correction term for the EE in non-Lorentz invariant critical points. However, the power-law correction term may not be universal in the sense that its exponent is not determined by the scaling dimension $\Delta$ of the relevant operator in the form of $L^{d+z+a-\Delta}$ and the exponent may change in different models. In our studies, the models have a dynamical exponent $z=2$, which is a strongly anisotropic critical point and can be described by the so-called conformal quantum critical point~\cite{qcp_z2, EE_cqcp}. The overlap of its ground state wave function with itself usually results in a partition function of a (1+1)D CFT. The scaling for the corner EE may be related to that in the (1+1)D CFT~\cite{EE_cqcp}. However, the correspondence between the relevant operator we added in the two-dimensional lattice model and operators in (1+1)D CFT can be complicated. Identifying the non-trivial correspondence may be helpful to explain the non-universal exponent in the power-law correction term. 

\section{Summary}
In this paper, we studied the bipartite EE induced by a small relevant deformation to Lorentz invariant quantum critical points in both one and two dimensions. We argue that, within the quantum critical region small relevant deformations at the critical point induce a universal power-law correction to the EE. The exponent for this term is determined by the scaling dimension of the relevant operator. We applied numerical simulations and scaling argument to verify the universal correction term. In one dimension, we used the finite entanglement scaling to study critical lattice models and confirm the universal power-law correction to the leading logarithmic scaling for the EE. In two dimensions, we studied the critical lattice free fermion models which is described by the Dirac fermion at the low energy. We utilized the correlation matrix method to calculate the EE. With different types of relevant deformations, the universal pow-law correction is verified and its exponent is also determined by the scaling dimension of the relevant operator. We also studied the corner contribution in the EE. The logarithmic contribution by the singularity point in the entangling surface is confirmed with two different scaling approaches. We also studied non-Lorentz invariant critical models with a dynamical exponent $z=2$. The pow-law correction for the EE induced by small relevant deformations is also found. However, such a power-law correction may not be universal. 

The models studied here in two dimensions are free fermion models. We expect the universal power-law correction to hold for non-trivial quantum critical points in interacting models, where there are anomalous correction to the dimension of relevant operators. In addition, there would be shift for the critical point at finite length scales. We expect there exists a sign change for the EE universal correction term at two different sides of the critical point. It is also interesting to further study EE in non-Lorentz invariant models at the special point $z=2$ from the (1+1)D CFT viewpoint. Another interesting topic is to study the correction for the EE for critical points close to the Lorentz invariant point $z \sim 1$. We leave these interesting topics for future studies.

\section{Acknowledgements}
We thank Hui-Ke Jin, Zi-Xiang Li and Long Zhang for helpful discussions. R.Z.H would like to thank Hai-Jun Liao and Tao Xiang for their kind hospitality during the recent visit to the Institute of Physics, Chinese Academy of Sciences, where part of this paper was completed. R.Z.H is supported by a postdoctoral fellowship from the Special Research Fund (BOF) of Ghent University. C.P is supported by the National Natural Science Foundation of China (Grant No. 12304182).

\nocite{*}
\bibliography{ref}

\begin{thebibliography}{42}
\bibitem [{not()}]{note_log}%
  \BibitemOpen
  \href@noop {} {\bibinfo {title} {Note that the logarithmic term remains invariant.}}\BibitemShut {Stop}%
\end{thebibliography}
\end{document}